\documentclass[twocolumn,showpacs]{revtex4}

\usepackage{amsmath}
\usepackage{amssymb}
\usepackage{epsfig}

\newcommand{\beq}{\begin{equation}} 
\newcommand{\eeq}{\end{equation}\noindent}
\newcommand{\bean}{\begin{eqnarray*}}
\newcommand{\eean}{\end{eqnarray*}\noindent}
\newcommand{\bea}{\begin{eqnarray}}
\newcommand{\eea}{\end{eqnarray}\noindent}

\newcommand{\cm}{{cm}}
\newcommand{\lab}{{lab}}

\begin{document}

\title{Neutrino-nucleus interaction rates at a low-energy beta-beam facility}

\author{Julien Serreau}
\email{serreau@thphys.uni-heidelberg.de} 
\affiliation{Institut f\"ur Theoretische Physik der Universit\"at Heidelberg, 
Philosophenweg 16, D-69120 Heidelberg, Germany}

\author{Cristina Volpe}
\email{volpe@ipno.in2p3.fr} 
\affiliation{Institut de Physique Nucl\'eaire, F-91406 Orsay Cedex, France}

\date{\today}

\begin{abstract}
We compute the neutrino detection rates to be expected at a low-energy beta-beam
facility. We consider various nuclei as neutrino detectors and compare the case 
of a small versus large storage ring.
\end{abstract}


\maketitle

\section{Introduction}

The pioneering experiment of R. Davis \cite{davis} has started the era of neutrino astronomy. 
Because they only have weak interactions with matter, neutrinos are precious messengers 
of what happens in the interior of stars, like our sun, or in explosive phenomena, such 
as Supernova type II explosions. Such astronomical neutrinos therefore provide an important 
source of information for our understanding of the life and death of stars. 
Nuclei are commonly used as detectors in neutrino observatories as well as in various 
experiments aiming at studying intrinsic neutrino properties, such as their masses 
and mixings.
A precise knowledge of neutrino-nucleus cross-sections is needed for the interpretation of 
these measurements and/or to study the feasibility of new projects. 
The understanding of neutrino-nucleus interactions is also of crucial importance for 
various astrophysical processes. 
Timely examples include neutrino nucleosynthesis \cite{haxton,woosley}, or 
the nucleosynthesis
of heavy elements during the so-called r-process \cite{bahanunucl,gail,bahaastro,qian,goriely}. 
If the latter takes place during the 
explosion of Supernovae type II, where a gigantic amount of energy is emitted as
neutrinos of all flavors, final abundances depend on several nuclear properties, among which
the interactions with neutrinos.

According to existing simulations, the average energy of neutrinos emitted from core-collapse
Supernovae is about $10$~MeV for electron neutrinos and about $20$~MeV for muon and tau 
neutrinos \cite{raffelt}. Notice however that, due to oscillations, electron
neutrinos can 
become hotter while traversing the star \cite{hax1999,smirnov,nuPb}. The predicted spectra cover 
the $50$~MeV region and present a tail up to about $100$~MeV \cite{raffelt}.
Reactor and solar neutrinos have typical energies in the $10$~MeV energy range, while 
accelerator and atmospheric neutrinos cover the GeV and multi-GeV range.
The various theoretical approaches
employed to describe neutrino-nucleus interactions therefore involve nuclear
as well as nucleonic degrees of freedom (for a review, see \cite{revuekubo,jpg}).
There are a number of open issues in this context. 
The A=2 system is the simplest case, for which the reaction cross sections 
can be estimated with high accuracy 
\cite{kubodera}. However, there is still an important quantity, 
namely L$_{1,A}$, related to the axial two-body current, which dominates the 
theoretical uncertainty in neutrino-deuteron interactions. For heavier nuclei, in 
the tens of MeV energy range, the reaction cross sections are dominated 
by collective modes, like the Gamow-Teller resonance or the Isobaric Analog State, 
which have been extensively studied in the past \cite{osterfeld}. As the neutrino 
impinging energy increases, transitions to states 
of higher multipolarity (such as the spin-dipole or higher forbidden
transitions) 
become important \cite{volpelead}. The latter also play an important role in 
the context of core-collapse Supernova physics \cite{kolbe,gail,jon,volpelead}. 
Although some information on these states
can be gathered through other probes, such as charge-exchange 
reactions \cite{osterfeld}, muon capture \cite{giai}, or inelastic electron
scattering \cite{electron}, the experimental information  is rather 
scarce.
Note that the understanding of neutrino-carbon reactions with 
neutrinos produced from the decay in flight of pions is still an 
open issue, for most of the theoretical calculations over-estimate the experimental 
value \cite{c12}.   
So far, measurements with low-energy neutrinos have been performed in a few cases
only, namely deuteron \cite{deut}, carbon \cite{expc12}, and iron \cite{iron}.
Systematic studies would be of great importance both for what concerns the 
interpolation from the MeV to the GeV neutrino energy range and the extrapolation 
to neutron-rich nuclei, as required in the astrophysical context.

Neutrino-nucleus interaction studies were one of the main physics issues of the
proposed ORLAND underground neutrino facility, which was based on a conventional
neutrino source (pion and muon decays) \cite{orland,jpg}.
A smaller version of the ORLAND project  is now under study
\cite{efremenko}.
At present, the
MINER$\nu$A project 
\cite{minerva} includes the study of neutrino-nucleus interactions for neutrino energies
in the GeV range. Here, we study the potential of a low-energy neutrino facility based 
on beta-beams, a novel method to produce neutrino beams \cite{zucchelli}.
This consists in boosting exotic ions which decay through beta-decay and produce 
pure, collimated and well-understood electron neutrino fluxes. Such a method could 
be exploited for a future facility at CERN \cite{zucchelli,mats}. High energy
beta-beams would be fired to a gigantic Cherenkov detector like UNO \cite{uno}, 
located in an (upgraded) Fr\'ejus underground laboratory to study, in particular, the possible 
existence of CP violation in the leptonic sector \cite{zucchelli,mats,mauro}.
The discovery potential with a very high $\gamma$ and a longer baseline is discussed
in \cite{jj,tmms}.

It has recently been proposed to use the beta-beam concept for the production of
low-energy neutrinos \cite{lownu}. Several laboratories will produce intense exotic beams in 
the near future and could, therefore, be possible sites for a low-energy beta-beam 
facility. These include GANIL, CERN, GSI, as well as the EURISOL project. 
Low-energy neutrino beams would offer an interesting opportunity 
to study various neutrino properties, such as e.g.\ the neutrino magnetic 
moment \cite{munu}, as well as neutrino-nucleus interactions, of interest for nuclear 
physics, particle physics and astrophysics. In the former case, one would exploit 
the ions at rest as an intense neutrino source, whereas, in the latter case, one would 
use boosted ions, which would be collected in a storage ring \cite{lownu}, as in the original high 
energy proposal. An important feature of such beta-beams is that the boost factor 
of the accelerated ions can be varied, allowing one to explore various neutrino 
energy ranges.

In this paper, we present for the first time charged-current neutrino-nucleus interaction 
rates achievable at a low-energy beta-beam facility. We consider two possible cases for
the dimensions of the storage ring, for which we inspire ourselves of the one planned 
in the future GSI facility \cite{gsi} and the one thought in the CERN baseline scenario 
\cite{zucchelli,mats}.
We consider various target nuclei as neutrino detectors, namely deuteron, oxygen, iron 
and lead, which are commonly used in existing or planned experiments 
\cite{orland}. Related work in the case of lead can be found in
\cite{gailnew}.

\section{Formalism}

\subsection{Neutrino fluxes and interaction rates}

The decay rate of a nucleus in the rest ($cm$) frame can be written as:
\beq
\label{e:1}
 \frac{dW}{dt}\Big|_\cm=\Phi_\cm (E_\nu)\,dE_\nu\,\frac{d^2\Omega}{4\pi}\,,
\eeq
where $E_{\nu}$ and $\Omega$ denote respectively the energy and the solid angle
of the emitted (anti-)neutrino, and where the neutrino flux $\Phi_\cm (E_\nu)$ is 
given by the well-known formula \cite{krane}:  
\beq
\label{e:2}
 \Phi_\cm (E_\nu)=b\,E_\nu^2\,E_e\,
 \sqrt{E_e^2-m_e^2}\, F(\pm Z,E_e)\,\Theta(E_e-m_e)\,.
\eeq
where the constant $b=\ln 2/m_e^5 ft_{1/2}$, with $m_e$ the electron mass
and $ft_{1/2}$ the ft-value. The quantities appearing in the above expression 
are the energy $E_e=Q-E_{\nu}$ of the emitted lepton (electron or positron), 
$Q$ being the $Q$--value of the reaction, and the Fermi function $F(\pm Z,E_e)$,
 which accounts for the Coulomb modification of the spectrum. 

In the laboratory frame, where the boosted nucleus has a velocity $v=\beta c$,
the decay rate reads:
\beq
 \frac{dW}{dt}\Big|_\lab=\frac{1}{\gamma}\,\Phi_\lab (E_\nu,\theta)\,
 dE_\nu\,\frac{d^2\Omega}{4\pi}\,,
\eeq
where $\gamma=1/\sqrt{1-\beta^2}$ is the time dilation factor
and where $E_{\nu}$ and $\Omega\equiv(\theta,\varphi)$ now denote the energy and
solid angle of the emitted (anti-)neutrino in the laboratory ($lab$) frame, 
$\theta$ 
being the angle of emission with respect to the beam axis. The boosted flux 
$\Phi_\lab (E_\nu,\theta)$ is given by:
\beq
\label{bflux}
 \Phi_\lab(E_\nu,\theta) = \frac{\Phi_\cm(E_\nu\gamma[1-\beta\cos\theta])}
 {\gamma[1-\beta\cos\theta]}\,.
\eeq

We consider a storage ring of total length $L$ with a straight sections of 
length $D$. In the stationary regime the mean number of ions 
in the storage ring is $\gamma\tau g$, where $\tau=t_{1/2}/\ln 2$ is the lifetime 
of the parent nuclei and $g$ is the number of injected ions per unit time. The 
total number of neutrinos emitted per unit time from a portion $d\ell$ of the 
decay ring is
\beq
 \frac{dN_\nu}{dt}=\gamma\tau\,g\times\frac{dW}{dt}\Big|_\lab
 \times\frac{d\ell}{L}\,.
\eeq

For simplicity, we consider a cylindrical detector of radius $R$ and depth $h$, 
aligned with one the straight sections of the storage ring, and placed at a distance 
$d$ from the latter. After integration over the useful decay path and over the volume 
of the detector, the total number of events per unit time is: 
\beq
\label{dNevdt}
 \frac{dN_{ev}}{dt}=g\tau nh\times
 \int_0^\infty dE_\nu\,\Phi_{tot}(E_\nu)\,\sigma(E_\nu)\,,
\eeq
where $n$ is the number of target nuclei per unit volume, $\sigma(E_\nu)$ is 
the relevant neutrino-nucleus interaction cross-section, and where
\beq
\label{phitot}
 \Phi_{tot}(E_\nu)=\int_0^D \frac{d\ell}{L}\,\int_0^h \frac{dz}{h}\,
 \int_0^{\bar{\theta}(\ell,z)} \frac{\sin\theta d\theta}{2}\,\Phi_{lab}(E_\nu,\theta)\,,
\eeq
with
\beq
\label{theta}
 \tan\bar{\theta}(\ell,z)=\frac{R}{d+\ell+z}\,.
\eeq

\subsection{Large versus Small Ring configurations}

\begin{figure*}[t]
\begin{center}
\epsfig{file=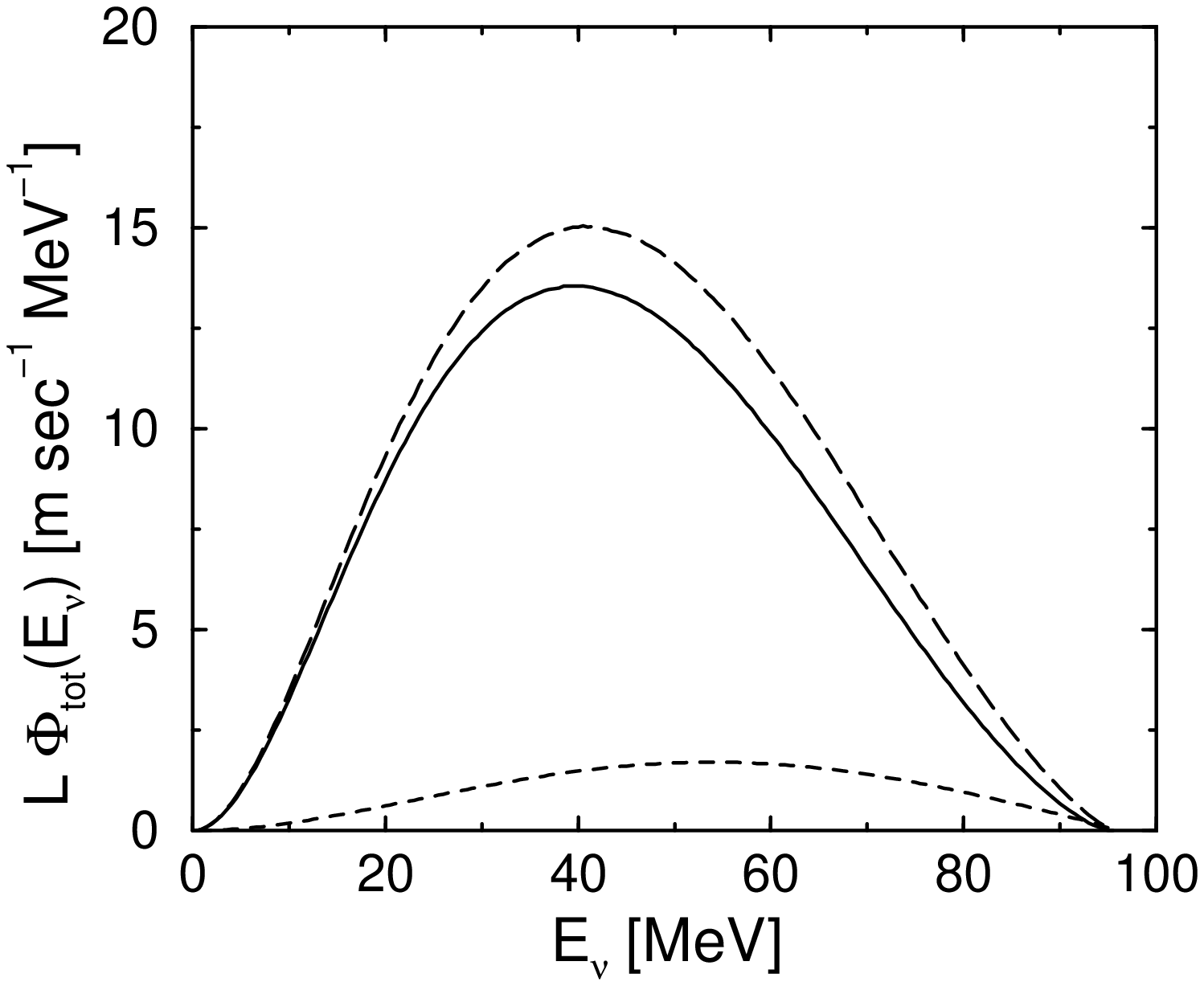,width=6.cm,angle=-90}
\hspace{2.cm}
\epsfig{file=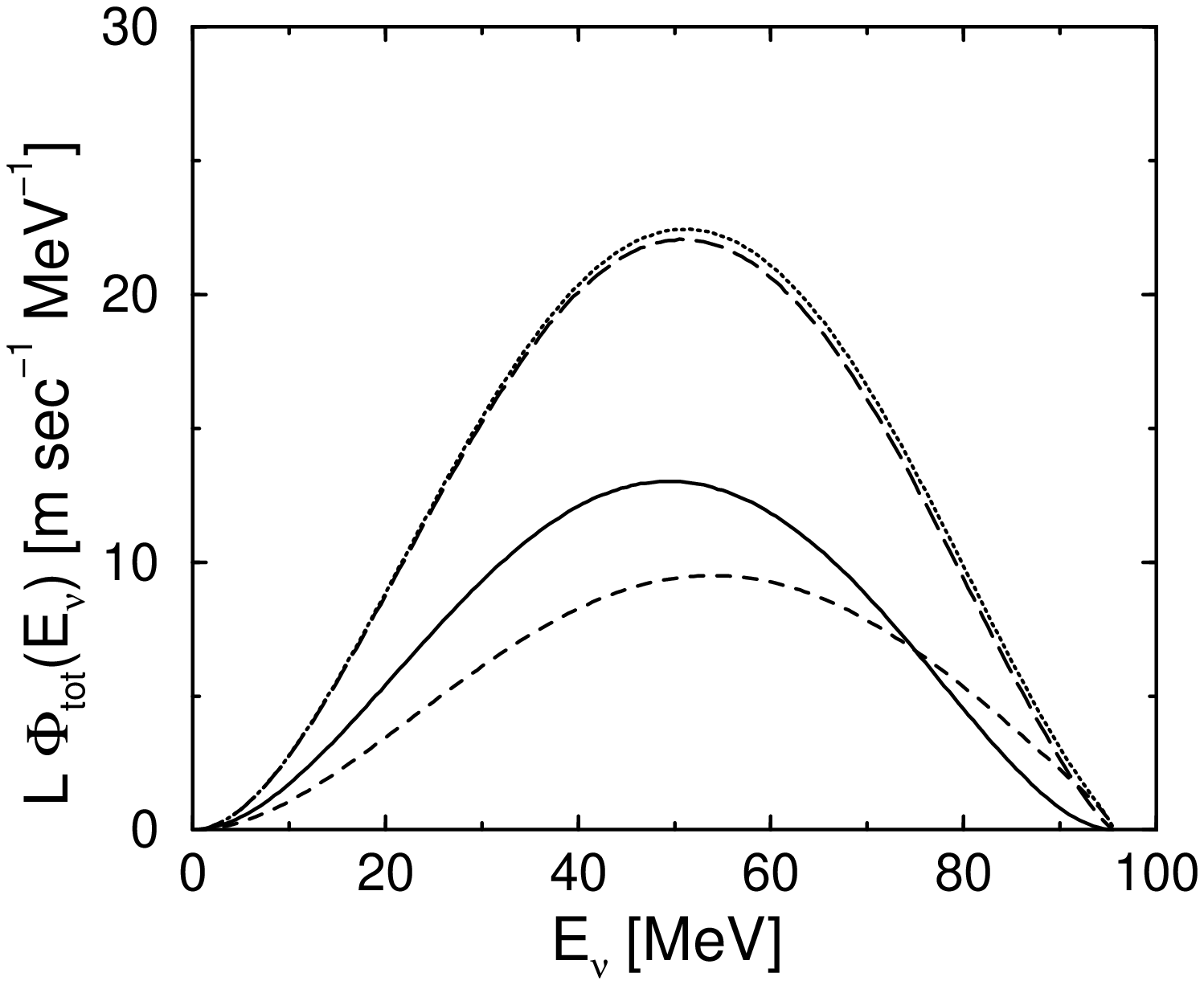,width=6.cm,angle=-90}
\end{center}
\caption{Neutrino fluxes scaled by the length of the storage ring 
 $L\Phi_{tot} (E_\nu)$: 
 The exact results obtained with Eqs.(\ref{phitot})-(\ref{theta})
 with a small storage ring $SR$ (solid lines) and a large 
 storage ring $LR$ (long-dashed lines) are shown. The left (right) figure shows the 
 fluxes impinging on the small (large) detector (the sizes are given in Tables \ref{tab:1}
 and \ref{tab:2}). For the small detector (left), the $LR$ result obtained with the 
 analytical estimate Eq.\ (\ref{LRSRflux}) coincide with the exact result and is not represented 
 here for clarity. For the large detector (right), it is also a very good approximation
 as shown by the dotted-line. The contribution $L\,G\,\Phi_{lab} (E_\nu,\theta=0)$
 from the RHS of Eq.\ (\ref{LRSRflux}) is also presented (dashed lines). 
 All fluxes are obtained with $^{18}$Ne boosted at 
 $\gamma=14$.} 
\label{fig:LRSR}
\end{figure*}

\noindent
The  storage ring geometry is characterized by the length of the straight sections $D$ 
and by its total length $L$. Below, we consider the cases of a small ($SR$) and a large ($LR$) 
ring configurations, characterized by ($D_{SR}$,$L_{SR}$) and ($D_{LR}$,$L_{LR}$) 
respectively. The results in both configurations can easily be related to one another 
by splitting the integral over the useful decay path $\int_0^{D_{LR}}\equiv 
\int_0^{D_{SR}} + \int_{D_{SR}}^{D_{LR}}$ in Eq.\ (\ref{phitot}). Up to trivial $1/L$
factors, the LHS corresponds to the $LR$ configuration and the first 
term on the RHS to the $SR$ configuration. The remaining integral can be given a 
simple analytical estimate if one can neglect the angular dependence of the flux 
under the integral. This happens when the angle under which the detector is seen 
from the extremity of the $SR$ decay path $\sim R/(d+D_{SR})$ is small compared
to $1/\gamma$, i.e. to the typical opening angle of the boosted flux.
In that case, we obtain, for the total flux (\ref{phitot}):
\beq
\label{LRSRflux}
 \Phi_{tot}^{LR} (E_\nu)\simeq \frac{L_{SR}}{L_{LR}}\times\Big\{
 \Phi_{tot}^{SR} (E_\nu) + G\,\Phi_{lab}(E_\nu,\theta=0)\Big\}\,,
\eeq
where the geometrical factor $G$ is given by:
\beq
 G=\frac{R^2}{4L_{SR}(d+D_{SR})}\left(1-\frac{d+D_{SR}}{d+D_{LR}}\right).
\eeq
The overall factor $L_{SR}/L_{LR}$ in (\ref{LRSRflux}) simply accounts for the
fact that the number of decaying ions per unit length is smaller in a larger 
storage ring, and the second term in brackets on the RHS represents the 
contribution from the longer useful straight section.
Figure \ref{fig:LRSR} shows a comparison between the exact 
flux obtained with Eqs.(\ref{phitot})-(\ref{theta}) in both  the $SR$ and the $LR$ 
configurations, and the analytic estimate Eq.\ (\ref{LRSRflux}), 
for the two possible
detector sizes considered in the following. We see that the analytical formula 
(\ref{LRSRflux}) works very well in the cases considered here. Besides, Figure 
\ref{fig:LRSR} shows that the contribution from the longer decay path only brings 
a $\sim10$~\% difference for the small detector and contributes a factor $\sim 2$ for
the larger detector. This already shows that the main difference between the 
$LR$ and $SR$ fluxes comes from the geometrical factor $L_{SR}/L_{LR}\simeq 1/15$.

Using the approximate formula for the total fluxes, we obtain an approximate 
relation between the total number of events in the $LR$ and $SR$ configurations:
\beq
\label{LRSRNev}
 \frac{dN_{ev}}{dt}\Big|_{LR}\simeq  \frac{L_{SR}}{L_{LR}}\times\left\{
 \frac{dN_{ev}}{dt}\Big|_{SR}+\gamma^2(1+\beta)^2 G\,gnh\,
 \langle\sigma\rangle_\gamma\right\},
\eeq
where $\langle\sigma\rangle_\gamma$ denotes the flux-averaged cross section
in the forward direction $\theta=0$:
\beq
\label{avcs1}
 \langle\sigma\rangle_\gamma=
 \frac{\int_0^\infty dE_\nu\,\Phi_{lab}(E_\nu,\theta=0)\,\sigma(E_\nu)}
      {\int_0^\infty dE_\nu\,\Phi_{lab}(E_\nu,\theta=0)}\,.
\eeq
Using Eq. (\ref{bflux}), the latter can be re-written as: 
\beq
\label{avcs2}
 \langle\sigma\rangle_\gamma=
 \frac{\int_0^\infty dE_\nu\,\Phi_{cm}(E_\nu)\,\sigma(\gamma(1+\beta)E_\nu)}
      {\int_0^\infty dE_\nu\,\Phi_{cm}(E_\nu)}\,.
\eeq
It is to be noted that, when the detector is placed close to the storage ring, as it is the case
here, the total rate (\ref{dNevdt}) depends non-trivially on the geometry of 
the latter.  For instance, as discussed above, we observe an approximate $1/L$ scaling at fixed 
$D/L$ in the small detector case. This is in contrast with the case of a far detector 
considered in the high energy beta-beam scenarios \cite{zucchelli,mauro,jj,tmms}, 
where the rate is simply proportional to the ratio $D/L$ of the straight 
section over the total length of the ring \footnote{For a distant detector ($d\gg L,D,h$), 
one has simply: $\Phi_{tot}(E\nu)\simeq\Phi_{lab}(E_\nu,\theta=0)\,\frac{D}{L}\,\frac{S}{4\pi d^2}$, 
where $S=\pi R^2$ is the transverse area of the detector. Similarly, one obtains, for the 
rate: $\frac{dN_{ev}}{dt}\simeq g\,\frac{D}{L}\,\frac{N_{target}}{4\pi d^2}
\, \gamma^2(1+\beta)^2\,\langle\sigma\rangle_\gamma$, where $N_{target}=n\pi R^2h$
is the total number of target nuclei.}. 

\section{Results}

\begin{figure}[t]
\begin{center}
\epsfig{file=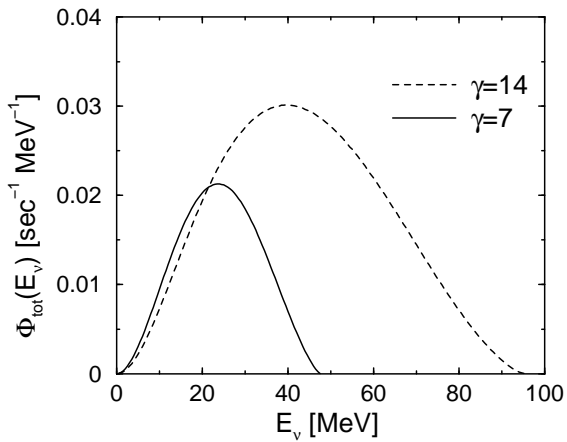,width=6.cm,angle=-90}
\end{center}
\caption{Neutrino fluxes $\Phi_{tot}(E_\nu)$ as a function of energy for $^{18}$Ne 
nuclei boosted at $\gamma=7$ and $\gamma=14$. This corresponds to the small ring
and small detector configuration.}
\label{fig:flux}
\end{figure}

Here, we present charged-current neutrino interaction rates with various target nuclei
as obtained from Eqs.\ (\ref{dNevdt})-(\ref{theta}) (Tables \ref{tab:1} and \ref{tab:2}).
Four possible nuclei are taken as typical examples,
namely deuteron, oxygen, iron and lead. 
A detailed study for the case of lead is also done in \cite{gailnew}.
The ``small ring'' we consider has 150 meter
straight sections and 450 meter total length, while the ``large ring''
has 2.5 km straight sections and 7 km total length. The detectors are located at
a distance 10 meters from the storage ring, to allow a maximum shielding of the
induced background in the ring \cite{matteo}. For the detector size we inspire
ourselves on the kinds considered for the proposed ORLAND facility
\cite{orland,avignone}.
The transverse size is chosen so as to catch as much as possible
of the boosted flux, which main contribution is concentrated in an opening
angle $\sim 1/\gamma$. More precisely, we choose as typical
dimensions ($R=$ radius, $h=$ depth): $R=1.5$~m
and $h=4.5$~m. We also consider the case of a large (kiloton-type) water detector
with $R=4.5$~m and $h=15$~m.
For all detectors here we assume a $100$~\% efficiency.
Finally, we have to specify the number of parent ions 
$g$ injected per unit time in the storage ring. According to the feasibility study 
\cite{mats}, $2\times10^{13}$ $^{6}$He/sec and $8\times10^{11}$ $^{18}$Ne/sec could be 
produced with an ISOLDE technique, giving about $g_{\bar{\nu}}=10^{13}$ $\bar{\nu}$/sec 
and $g_\nu=5\times 10^{11}$ $\nu$/sec respectively \cite{mats}.
An important feature of beta-beams is that the number and average 
energy of neutrinos entering the detector depend on the boost factor
$\gamma$ of the parent ion, which can be varied. We present results for 
two different values, namely $\gamma=7$ (Table I) and $\gamma=14$ (Table II). 
The corresponding neutrino fluxes are presented in Figure \ref{fig:flux} and 
range up to about 50 and 100 MeV respectively. 

\begin{figure}[t]
\begin{center}
\epsfig{file=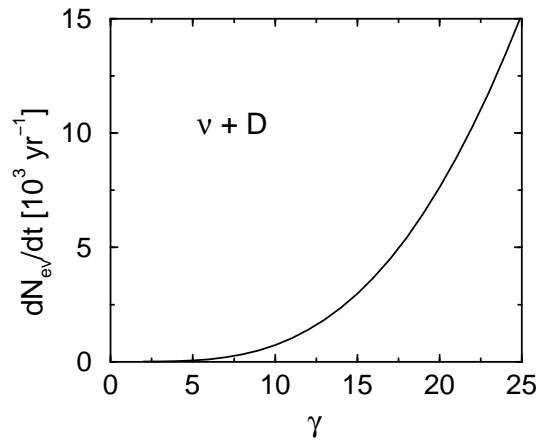,width=6.cm,angle=-90}
\end{center}
\caption{The total rate for the reaction $\nu + {\rm D}$ as a function of the
boost factor $\gamma$. This corresponds to the small ring and small detector 
configuration.} 
\label{fig:rate}
\end{figure}

Let us discuss the number of events shown in Tables \ref{tab:1} and \ref{tab:2}.
The differences between the $\nu$--induced versus $\bar{\nu}$--induced reactions
is a combined effect of the relative intensities $g_\nu/g_{\bar{\nu}}=1/20$ 
and of the different interaction cross-sections: 
the ratio $\sigma(\nu+{\rm D})/\sigma(\bar{\nu}+{\rm D})$ is roughly $2$ in the whole energy 
range considered here \cite{revuekubo}; from \cite{nuO}, one can see that 
$\sigma(\nu+{\rm ^{16}O})/\sigma(\bar{\nu}+{\rm ^{16}O})$ is about $0.5$ on average in 
the energy range relevant to the case $\gamma=7$, namely $20~{\rm MeV} \lesssim E_\nu 
\lesssim 40~{\rm MeV}$, and about $1.5$ on average in the range $40~{\rm MeV} \lesssim 
E_\nu \lesssim 80~{\rm MeV}$, relevant for the case $\gamma=14$. The very low rates 
obtained for oxygen with $\gamma=7$ despite the large detector size are due to the
$15$~MeV threshold in the interaction cross-section. Next, we observe that the suppression of the 
rates in the $LR$ configuration as compared to the $SR$ case for a given $\gamma$ roughly 
corresponds to the geometrical factor $L_{SR}/L_{LR}$, as expected from the previous discussion. 
In fact, the difference between the $LR$ and $SR$ rates can be fully understood by means of
the approximate relation Eq.\ (\ref{LRSRNev}). This formula can be used to rescale our 
results for other possible dimensions of the storage ring. To this aim, we give the relevant 
values of $\langle \sigma \rangle_\gamma$ in each case. 
When going from $\gamma=7$ to $\gamma=14$, the neutrino fluxes 
become more collimated and the typical energy of the neutrinos increases. This, together with 
the fact that the neutrino-nucleus interaction cross sections rapidly rise with the impinging 
neutrino energy, increases the number of events by more than an order of magnitude. Figure 
\ref{fig:rate} illustrates the rapid rise of the total rates with increasing $\gamma$. 
Note that, in the present case, where the detector is relatively close to the storage ring,
the total rates do not have a simple scaling with the detector size, due to the non-trivial 
angular dependence of the impinging neutrino flux. 

\begin{table*}
 \begin{tabular}{|ccc|ccc|ccc|ccc|ccc|ccc|}
 \hline
 &                  &&&                  &&&       &&&                               &&&            &&&            &\\
 &Reaction          &&& Ref.             &&& Mass  &&& $\langle\sigma\rangle_\gamma$ &&& Small Ring &&& Large Ring &\\ 
 &                  &&&                  &&&       &&&                               &&&            &&&            &\\
 \hline
 &$\nu +$D          &&&\cite{revuekubo}  &&&  35   &&& 36.30                         &&&  194       &&&  14        &\\
 &$\bar\nu +$D      &&&\cite{revuekubo}  &&&  35   &&& 23.16                         &&&  2494      &&&  178       &\\
 &$\nu + ^{16}$O    &&&\cite{nuO}        &&&  952  &&& 3.33                          &&&  60        &&&  6         &\\
 &$\bar\nu +^{16}$O &&&\cite{nuO}        &&&  952  &&& 5.04                          &&&  2125      &&&  192       &\\
 &$\nu +^{56}$Fe    &&&\cite{nuFe}       &&&  250  &&& 137.86                        &&&  872       &&&  63        &\\
 &$\nu +^{208}$Pb   &&&\cite{volpelead}  &&&  360  &&& 2931.24                       &&&  7598      &&&  545       &\\ 
 \hline
 \end{tabular}
 \caption{Number of events per year   for $\gamma=7$ in the small ($L_{SR}=450$~m,
 $D_{SR}=150$~m) and large ($L_{LR}=7$~km, $D_{LR}=2.5$~km) ring configurations.
 These results are obtained by using the exact formulas of Eqs.(6-\ref{theta}).
 The detector is located at $d=10$~m away from the ring and has
 dimensions
 $R=1.5$~m and $h=4.5$~m for the D (D$_2$O), $^{56}$Fe and $^{208}$Pb, and $R=4.5$~m
 and $h=15$~m for the case of $^{16}$O (H$_2$O), where $R$ is the radius and
 $h$ is the depth of the detector. The corresponding masses are given in tons.
 The results in the large ring configuration can be precisely understood from those 
 in the small ring configuration by means of the analytical formula Eq.\ (\ref{LRSRNev}).
 We give the flux-averaged cross section in the forward direction
 $\langle\sigma\rangle_\gamma$ (see Eqs.\ (\ref{avcs1})-(\ref{avcs2})) in units of 
 $10^{-42}$~cm$^2$. The latter can be used to rescale the present rates for different
 sizes of the storage ring using Eq.\ (\ref{LRSRNev}).
 The relevant cross-sections are taken from the indicated references. 
 The results are obtained with $1$~year =$~3.2 \times 10^{7}$~s.}
\label{tab:1}
\end{table*}

\begin{table*}
 \begin{tabular}{|ccc|ccc|ccc|ccc|ccc|ccc|}
 \hline
 &                  &&&                  &&&       &&&                               &&&            &&&            &\\
 &Reaction          &&&  Ref.            &&& Mass  &&& $\langle\sigma\rangle_\gamma$ &&& Small Ring &&& Large Ring &\\ 
 &                  &&&                  &&&       &&&                               &&&            &&&            &\\
 \hline
 &$\nu +$D          &&&\cite{revuekubo}  &&&  35   &&& 184.47                        &&&  2363      &&&  180       &\\
 &$\bar\nu +$D      &&&\cite{revuekubo}  &&&  35   &&& 96.03                         &&&  25779     &&&  1956      &\\
 &$\nu + ^{16}$O    &&&\cite{nuO}        &&&  952  &&& 174.28                        &&&  6054      &&&  734       &\\
 &$\bar\nu +^{16}$O &&&\cite{nuO}        &&&  952  &&& 102.00                        &&&  82645     &&&  9453      &\\
 &$\nu +^{56}$Fe    &&&\cite{nuFe}       &&&  250  &&& 1402.11                       &&&  20768     &&&  1611      &\\
 &$\nu +^{208}$Pb   &&&\cite{volpelead}  &&&  360  &&& 16310.16                      &&&  103707    &&&  7922      &\\ 
 \hline
 \end{tabular}
 \caption{Same as Table \ref{tab:1} for $\gamma=14$.}
\label{tab:2}
\end{table*}

It is important to emphasize the complementarity between low-energy beta-beams
and conventional neutrino facilities \cite{orland}. The latter provide intense 
sources of electron and muon neutrinos and cover the very low energy region, similar
to the case $\gamma=7$ for the beta-beam. Let us mention that for comparable neutrino 
intensities, the rates presented in Table \ref{tab:1} are comparable to those obtained
with conventional schemes with detectors located at about $50$~meters from the 
source.  
Low-energy beta-beams would produce pure electron neutrino beams and, by varying the
boost factor $\gamma$, would offer a unique opportunity to study neutrino-nucleus 
interactions over a wide range of energies.

To conclude, the present study demonstrates that, with typical parameters available 
from existing studies \cite{mats}, significant interaction rates can be achieved at a 
low-energy beta-beam facility. A small ring -- with as long as possible straight 
sections -- is the preferred configuration in the case of a close detector.
The rates raise rapidly with increasing $\gamma$. We think our results 
are encouraging and we hope they will trigger further investigations, including, in 
particular, detailed simulations of the detectors.

\vspace{.2cm}
We thank J. Bouchez and M. Magistris for useful discussions,
R. Lombard and M. Mezzetto for careful reading of
the manuscript.

\end{document}